# Diffusion-Based Probabilistic Modeling for Hourly Streamflow Prediction and Assimilation

Wencong Yang[1], Haoyu Ji[1], Leo Lonzarich[1], Yalan Song[1], Chaopeng Shen[1,*]

[1] Civil and Environmental Engineering, The Pennsylvania State University, University Park, PA 16802

* Corresponding author. cshen@engr.psu.edu

**Abstract**

Hourly predictions are critical for issuing flood warnings as the flood peaks on the hourly scale can be distinctly higher than the corresponding daily ones. Currently a popular hourly data-driven prediction scheme is multi-time-scale long short-term memory (MTS-LSTM), yet such models face challenges in probabilistic forecasts or integrating observations when available. Diffusion artificial intelligence (AI) models represent a promising method to predict high-resolution information, e.g., hourly streamflow. Here we develop a denoising diffusion probabilistic model (h-Diffusion) for hourly streamflow prediction that conditions on either observed or simulated daily discharge from hydrologic models to generate hourly hydrographs. The model is benchmarked on the CAMELS hourly dataset against record-holding MTS-LSTM and multi-frequency LSTM (MF-LSTM) baselines. Results show that h-Diffusion outperforms baselines in terms of general performance and extreme metrics. Furthermore, the h-Diffusion model can utilize the inpainting technique and recent observations to accomplish data assimilation that largely improves flood forecasting performance. These advances can greatly reduce flood forecasting uncertainty and provide a unified probabilistic framework for downscaling, prediction, and data assimilation at the hourly scale, representing risks where daily models cannot.

1. Introduction

Hourly flood warnings must grapple with flash floods—rapid, high‑magnitude rises in discharge that appear and subside within hours. Flash floods account for roughly 85 % of flood events and cause more than 5 000 deaths worldwide each year (World Meteorological Organization, 2020). Their short (<6 h) rise times and high peak discharges produce the highest mortality rate among hydrologic disasters (Georgakakos et al., 2022) and can even divert rivers from their channels. Historical examples include Hurricane Mitch in 1998, where flash floods and landslides killed over 11 000 people in Central America (Hellin et al., 1999). In the United States, especially Texas, flash floods occur when intense rainfall exceeds soil infiltration, causing water levels to rise within minutes; the short warning times make them the most dangerous type of flood (Furl et al., 2018). Flash floods kill more people in the United States each year than tornadoes or hurricanes (National Severe Storms Laboratory, n.d.). Crucially, the difference between the daily-averaged discharge and the disaster-inducing hourly peak discharge can be substantial. Therefore, accurate models for hourly discharge are of great importance.

Data-driven models such as long short-term memory (LSTM) networks have achieved state-of-the-art performance in rainfall–runoff modelling, but standard LSTMs, however, are typically trained at daily resolution. Extending them to hourly predictions demands sequences of 8 760 time steps per year, leading to prohibitive computational cost and difficulty learning long sequences. To address this, multi-timescale LSTM (MTS-LSTM) architectures (Gauch et al., 2021) process older inputs at coarse temporal resolution and recent inputs at fine resolution. showed that MTS-LSTM captures sub-daily dynamics, reduces computational demand and outperforms the NOAA National Water Model. Recently, multi-frequency LSTM (MF-LSTM, Acuña Espinoza et al., 2025) removes this separation by allowing different temporal frequencies within a single LSTM cell and permitting distinct input dimensions, achieving higher computational efficiency.

Generative diffusion models, originally developed for image synthesis, are emerging as powerful tools for super-resolution tasks, which produce higher spatiotemporal resolution of outputs than the inputs. Unlike parametric approaches, diffusion models iteratively denoise random signals to sample from complex distributions without presupposing a runoff distribution. Ou et al. (2025) used a diffusion-based runoff model (DRUM) to generate probabilistic flood forecasts and demonstrated that it outperforms benchmark methods for extreme floods and extends early warning lead times, but the simulation is on a daily scale. Diffusion models naturally produce fine-resolution outputs conditioned on coarser inputs. For precipitation, the Wavelet Diffusion Model learns precipitation structures in the wavelet domain and achieves a ten-fold spatial downscaling from 10 km to 1 km while delivering a nine-fold inference speedup compared to pixel-based diffusion models (Yi et al., 2025). PrecipDiff downscales satellite precipitation from 10 km to 1 km and reduces bias using residual learning based solely on precipitation data (Dai & Ushijima-Mwesigwa, 2025). In wind forecasting, WindSR integrates sparse observations with simulations through a dynamic-radius data-assimilation scheme, producing super-resolved wind fields and reducing bias by about 20 % relative to CNN and GAN baselines (Ma et al., 2025). These successes illustrate that diffusion AI can deliver fine-grained geophysical fields and incorporate observations via inpainting without extensive training datasets. Given their probabilistic nature and ability to condition on coarse inputs and assimilate data, diffusion models are a natural candidate for hourly streamflow prediction.

Here we develop a denoising diffusion probabilistic model (h-Diffusion) that conditions on coarse discharge and rainfall to generate full hourly hydrographs. We also propose an efficient training strategy that alternates between daily and hourly updates. We benchmark their performance on the CAMELS hourly dataset by comparing them against record-holding MTS-LSTM and MF-LSTM baselines. Uniquely to diffusion models, we demonstrate that (a) diffusion-based inpainting can assimilate recent observations without additional training to improve flood forecasts; and (b) diffusion can produce probabilistic forecasting without an explicit ensemble (Ou et al., 2025). The core questions guiding this manuscript are:

(1) Can h-Diffusion achieve state-of-the-art performance (e.g., MTS-LSTM, MF-LSTM), either for downscaling daily observed discharge or making predictions?

(2) To what extent can diffusion inpainting exploit sparse observations to boost forecast skill?

## 2. Data and Methods

### 2.1 Data

To enable direct comparison with previous studies (Acuña Espinoza et al., 2025; Gauch et al., 2021), we use the same dataset, which includes 516 basins from the CAMELS-US database (Addor et al., 2017). The dataset provides hourly streamflow observations retrieved from the USGS Water Information System, hourly meteorological forcings from the North American Land Data Assimilation System (NLDAS, Xia et al., 2012), daily forcings from Daymet and Maurer (Addor et al., 2017), and 26 selected static catchment attributes. The full list of variables can be seen in Gauch et al. (2021).

Following the experiments in Gauch et al. (2021) and Acuña Espinoza et al. (2025), all models were trained with data from 1990-10-01 to 2003-09-30 and tested from 2003-10-01 to 2014-09-30. The exact testing period varies by forcing product because of data availability: Daymet is available until 2014-12-31 and Maurer is available only until 2008-12-31 in CAMELS-US. The detailed testing periods for each experiment are described in the corresponding method sections.

### 2.2 Diffusion-based Hourly Streamflow Models

We developed h-Diffusion, a conditional diffusion model that generates hourly streamflow values $q_h$ within a 72-hour window. The model learns the complex probability distribution of streamflow and enables statistical sampling from it.

**Diffusion model**

Diffusion models learn complex, nonlinear data distributions and generate diverse, physically consistent samples conditioned on hydrologic inputs. We formulate the prediction of hourly streamflow using a conditional diffusion model as:

$$q_h = f(F_d, F_h, A, q_{ref}) \qquad (1)$$

where $F_d$ is daily forcing data, $F_h$ is hourly forcing data, $A$ is catchment attributes, and $q_{ref}$ is a reference daily streamflow value at the prediction window.

A diffusion model consists of two processes: a forward (noising) process and a backward (denoising) process. In the forward process, each observed streamflow sample $q_{h,obs}^0$ is gradually perturbed by Gaussian noise $\epsilon \sim N(0, 1)$ over $T = 1000$ diffusion steps, following the denoising diffusion probabilistic model (DDPM, Ho et al., 2020) framework. The process is defined as:

$$q_h^{(t)} = \sqrt{1 - \beta_t} q_h^{(t-1)} + \sqrt{\beta_t} \epsilon, \epsilon \sim N(0, 1) \qquad (2)$$

where $t \in [1, T]$ is the diffusion step and $\beta_t$ is the noise-variance schedule that controls how much Gaussian noise is added at each step. After $T$ steps, the data distribution becomes approximately standard normal.

In the reverse process, we start from Gaussian noise $q_h^{(T)}$ and iteratively recover cleaner samples $q_h^{(t-1)}$ using:

$$p\left(q_h^{(t-1)} | q_h^{(t)}\right) = N\left(q_h^{(t-1)}; \frac{1}{\sqrt{1-\beta_t}}\left(q_h^{(t)} - \beta_t \epsilon_\theta\left(q_h^{(t)}, t, c\right)\right), \sigma_t^2 I\right) \qquad (3)$$

where $\epsilon_\theta\left(q_h^{(t)}, t, c\right)$ is a neural network that predicts the noise component to be removed, and $c$ denotes conditional information.

In practice, the network is trained to minimize the expected mean-squared error between the true noise and the predicted noise:

$$L_\theta = E\left[\left\|\epsilon - \epsilon_\theta\left(q_h^{(t)}, t, c\right)\right\|^2\right] \qquad (4)$$

During inference, the reverse denoising (Equation 3) runs for 50 steps following the Denoising Diffusion Implicit Model (DDIM, J. Song et al., 2022) scheduler, which preserves sample quality while reducing computational cost. The final clean sample is the predicted 72-hour sequence of hourly streamflow.

**Network architecture**

The backbone of the diffusion model, i.e., the noise-prediction network $\epsilon_\theta$, inherits the Multi-Timescale LSTM (MTS-LSTM) architecture from Gauch et al. (2021). It consists of a daily-scale LSTM branch that runs for 362 days to warm up hidden states and an hourly-scale LSTM branch that models the 3-day (72-hour) high-frequency dynamics. This architecture captures multi-scale dependencies while avoiding the computational burden of long sequential modeling. As shown in Figure 1(a), in this network, daily forcing data $F_d$ and catchment attributes $A$ are input to the daily LSTM to produce the last-step hidden and cell states ($h_d$, $c_d$). They are then transformed through a multilayer perceptron (MLP) into ($h_h$, $c_h$), the initial states for the hourly branch. Unlike the original MTS-LSTM, the hourly branch here does not directly predict streamflow. Instead, it takes the current noisy sample $q_h^{(t)}$, timestep $t$, hourly forcing $F_h$, catchment attributes $A$, and a conditional reference daily streamflow $q_{ref}$ to estimate the noise term $\epsilon$ to be removed from $q_h^{(t)}$.

**Conditional input and mixed-sample training**

To provide daily hydrologic context, we include a reference daily streamflow $q_{ref}$ at the same 72-hour window as a conditioning variable. As illustrated in Figure 1(a), two options are used for $q_{ref}$: (1) the observed daily streamflow $q_{d,obs}$, representing a downscaling mode to refine daily observations into hourly timeseries; and (2) the simulated daily streamflow $q_{d,sim}$ from a hydrologic model, representing a prediction mode to correct simulated biases and also downscale the simulation. A flag variable indicates which mode the model is in. Training samples are mixed: for each 72-hour window, both observed- and simulated-reference samples are included. This strategy has three advantages: (1) strong contextual information from observed daily data helps the model focus on hourly dynamics; (2) simulated daily values provide physically constrained priors for ungauged or prediction periods; (3) it allows flexible applications, i.e., downscaling observed daily flow ($q_{d,obs}$, $flag = 0$), downscaling simulations from any daily model ($q_{d,sim}$, $flag = 0$), or full prediction ($q_{d,sim}$, $flag = 1$).

Similar to Gauch et al. (2021), our model also supports multi-forcing input. In addition to daily and hourly meteorological forcings ($F_d$ and $F_h$), the reference daily streamflow $q_{d,sim}$ can also be a multi-dimensional sequence derived from different forcing products (e.g., Daymet, NLDAS,

Maurer). To ensure compatibility between the one-dimensional observed reference ($q_{d,obs}$) and the multi-dimensional simulated reference ($q_{d,sim}$), a one-dimensional convolutional neural network (1D-CNN) layer is applied to map $q_{d,sim}$ to a unified one-dimensional latent space before inputting into the noise-prediction network $\epsilon_\theta$ as conditional information.

**Model inference**

As illustrated in Figure 1(b), the diffusion-based hourly streamflow model, h-Diffusion, produces hourly streamflow estimates by progressively denoising an initial Gaussian noise $q_h^{(T)}$. In this study, we focus on predictive skill of the model, and therefore, only simulated daily streamflow $q_{d,sim}$ is used as the daily reference input during testing. This ensures our results are comparable with previous study (Acuña Espinoza et al., 2025; Gauch et al., 2021). Although the model generates 72-hour window streamflow, we only use values from the last 24 hours of the window for testing. In summary, the h-Diffusion predictions can be expressed as:

$$q_h = f\left(F_d,\ F_h,\ A,\ q_{d,sim},\ flag = 1\right) \quad (5)$$

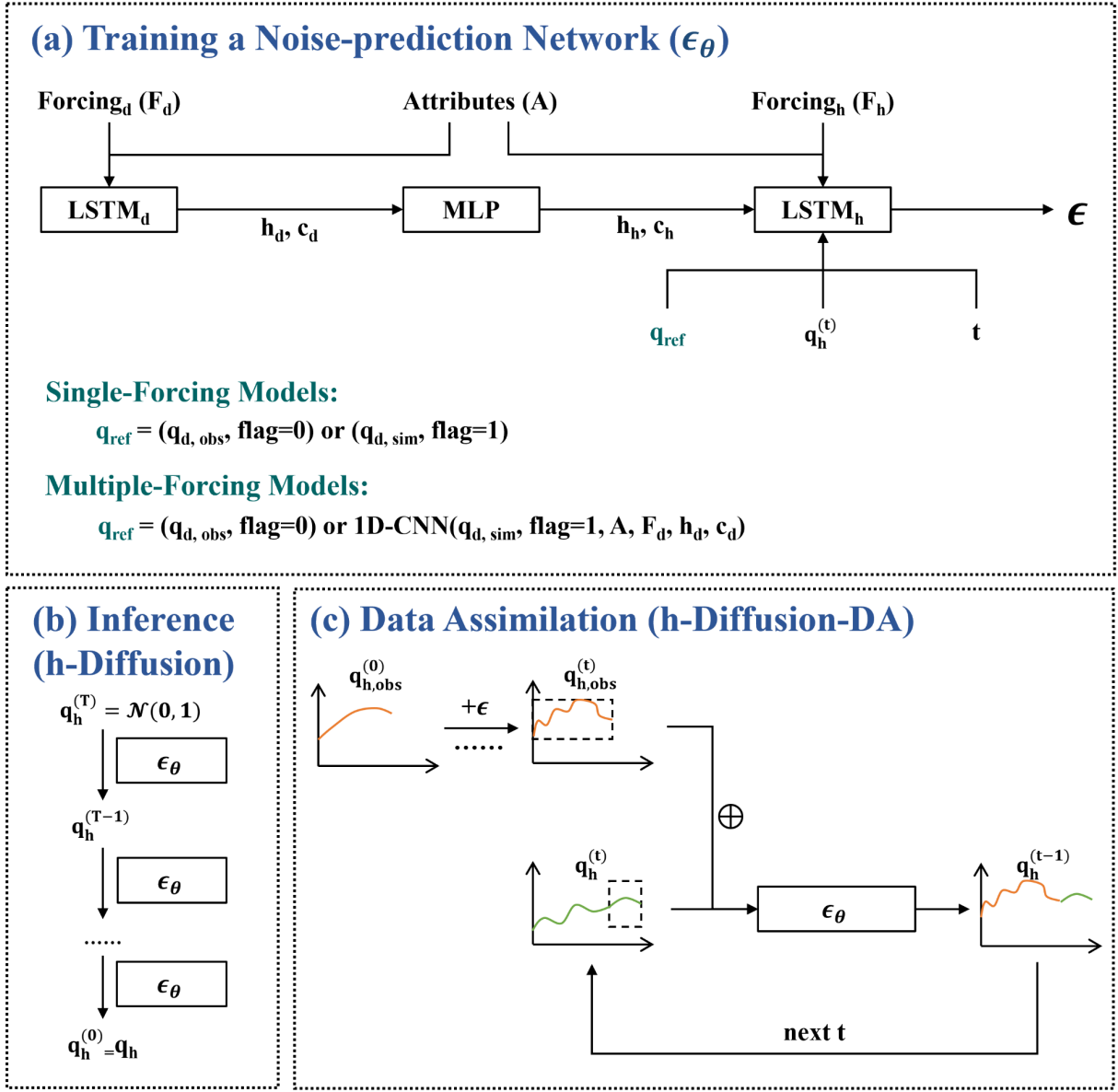

*Figure 1. Diffusion based hourly streamflow prediction model. $q_{d,\,obs}$: daily streamflow observations; $q_{d,\,sim}$: daily streamflow simulations from the hydrologic models; $h_d$, $c_d$: final hidden states of the daily LSTM branch; $h_h$, $c_h$: initial hidden states of the daily LSTM branch; $q_{h,\,obs}$: hourly streamflow observations; $q_h$: hourly streamflow predictions from the diffusion model. $t$: noising or denoising step. $\epsilon$: diffusion noise.*

2.3 Inpainting for Data Assimilation

One of the key advantages of diffusion models is their generic data assimilation capability, i.e., the inpainting technique in the area of generative artificial intelligence (Lugmayr et al., 2022).

By injecting a few known observations into the denoising process, the diffusion model can reconstruct complete samples that remain statistically consistent with both the learned distribution and the given observations. Unlike data integration approaches (Fang & Shen, 2020; Feng et al., 2020), inpainting does not need systematic training data; Unlike data assimilation approaches (Nearing et al., 2022a), inpainting does not solve an optimization problem and is thus more efficient. We adopt the RePaint algorithm (Lugmayr et al., 2022) to augment our h-Diffusion model into h-Diffusion-DA, which integrates streamflow observations directly during inference, as shown in Figure 1(c). In this study, we focus on predicting the streamflow of the last 24 hours within each 72-hour window. We assume that a few recent streamflow observations are available several hours before this target period. Specifically, the last 5 hours before the 24-hour prediction are known. According to the RePaint algorithm, at each denoising step $t$, the diffusion model generates a standard sample for the unknown (to-be-predicted) part:

$$q_{unknown}^{(t-1)} = \frac{1}{\sqrt{1-\beta_t}}\left(q_{unknown}^{(t)} - \beta_t \epsilon_\theta\left(q_{unknown}^{(t)}, t, c\right)\right) + \sigma_t z, \; z \sim N(0, I) \quad (6)$$

To align the known (observed) part with the same noise level, Gaussian noise is added to the observations:

$$q_{known}^{(t-1)} = \sqrt{1-\beta_t} q_{known}^{(0)} + \sqrt{\beta_t} \epsilon, \; \epsilon \sim N(0, 1) \quad (7)$$

Then, the updated sample at step $t-1$ is created by replacing the unknown part with the known part:

$$q^{(t-1)} = M \odot q_{known}^{(t-1)} + (1-M) \odot q_{unknown}^{(t-1)} \quad (8)$$

where $M$ is a binary mask indicating the window positions of available observations, and $\odot$ is elementwise multiplication.

A direct replacement (Equation 8) may produce discontinuities between known and unknown sequences. The RePaint algorithm introduces a harmonization step that periodically re-noises and re-denoises the sequence to smooth inconsistencies near the observation boundaries across the reverse process. Specifically, a re-noising step adds a small amount of noise again at a jump position, which occurs after every 5 normal denoising steps in our case. At each jump position, the model repeats the re-noising–denoising cycle 5 times, gradually refining the sample.

Unlike conventional hourly streamflow models, such as MTS-LSTM, MF-LSTM, or process-based hydrologic models, diffusion models with inpainting do not require an independent data assimilation framework (e.g., variational data assimilation or ensemble Kalman filter) to ingest observations. Inpainting is an intrinsic component of the model's inference stage.

Moreover, diffusion-based models differ from traditional data-integration models (Feng et al., 2020) that require explicit past observations during training. Once trained, h-Diffusion-DA can be used directly for either pure prediction or observation assimilation, which is highly flexible for applications across gauged and ungauged catchments.

2.4 Experiment Setup

This study compares the hourly streamflow prediction performance of four methods across 516 CAMELS-US basins: MTS-LSTM, MF-LSTM, h-Diffusion, and h-Diffusion-DA. Both h-Diffusion and h-Diffusion-DA require daily streamflow simulations ($q_{d,sim}$) from a physical hydrologic model. Here we employ the differentiable Hydrologiska Byråns Vattenbalansavdelning ($\delta HBV$) model, which provides high simulation performance and computational efficiency across CAMELS-US, and also supports predictions in ungauged basins through parameter learning (Feng et al., 2022). All hourly models and the dHBV model were trained using data from 1990-10-01 to 2003-09-30. Since the original training periods of MTS-LSTM and MF-LSTM remain unchanged, their pretrained models were reused for comparison.

Testing periods differ depending on the forcing products. Each hourly model was tested under two forcing-input configurations. (1) In single-forcing configuration, each model uses a single forcing product. The hourly models (MTS-LSTM, MF-LSTM, h-Diffusion, and h-Diffusion-DA) are driven only by the NLDAS hourly forcing, the only available hourly meteorological dataset in this study. Meanwhile, the daily $\delta HBV$ that provides the conditional streamflow simulations is driven solely by the Daymet daily forcing, which achieves superior performance among the available daily products (Feng et al., 2023). (2) In multiple-forcing configuration, models used three daily forcing datasets (Daymet, Maurer, and temporally aggregated NLDAS) as daily inputs $F_d$. For hourly inputs $F_h$, disaggregated Daymet (daily values repeated for 24 hours), disaggregated Maurer, and original NLDAS data were used together. Correspondingly, daily dHBV simulations driven by Daymet, Maurer, and NLDAS served as conditional inputs for h-Diffusion and h-Diffusion-DA. The testing period for single-forcing models was set to 2008-10-01 to 2014-09-30, and for multiple-forcing models it was 2003-10-01 to 2008-09-30.

We evaluated model performance using the Nash–Sutcliffe Efficiency (NSE, Nash & Sutcliffe, 1970) and the Continuous Ranked Probability Score (CRPS, Gneiting & Raftery, 2007). NSE assesses overall performance for hourly streamflow across the full flow range. CRPS is a negatively oriented score (like RMSE) that measures the distance between the predicted cumulative probability distribution and the observed outcome (Gneiting & Raftery, 2007; Ou et al., 2025). It jointly reflects prediction accuracy and the tightness of uncertainty—penalizing both large errors and overly broad or overconfident probabilistic spreads, with 0 being the best match:

$$CRPS = \sum_{i=1}^{n} P(q_i)^2 \cdot I(q_i \leq q_{obs}) + \sum_{i=1}^{n} \left(P(q_i) - 1\right)^2 \cdot I(q_i > q_{obs}) \qquad (9)$$

where $P(q_i)$ is the cumulative distribution function of the prediction $q_i$, $n$ is the number of ensemble realizations, and $I$ is the indicator function. Lower CRPS values correspond to better predictions.

Since h-Diffusion and h-Diffusion-DA are stochastic, we generated 10 realizations of hourly streamflow and used the ensemble mean for evaluation. Similarly, the original MTS-LSTM and MF-LSTM models produced ensemble predictions from 10 independently trained models with different random seeds, which also enabled ensemble mean evaluation. The average CRPS was calculated for all hourly streamflow values per catchment. To further assess the predictive skill across different levels of streamflow, the average CRPS was computed for streamflow values above selected percentiles (top 10%, 5%, 2%, 1%, and 0.5%). Since CRPS has physical units (mm h$^{-1}$) and its value depends on the streamflow magnitude of each basin, it is not suitable for direct comparison across catchments or models. To allow comparison, we further compute the Continuous Ranked Probability Skill Score (CRPSS, Alfieri et al., 2014), which evaluates model performance relative to a reference prediction:

$$CRPSS = 1 - \frac{\overline{CRPS}}{\overline{CRPS}_{ref}} \qquad (10)$$

The CRPSS ranges in $[-\infty, 1]$, where CRPSS>0 indicates improvement over the reference prediction, and higher CRPSS values represent greater gains in predictive skill relative to the reference model. In this study, we used the average CRPSs of MTS-LSTM and MF-LSTM as two references.

## 3. Results and Discussions

### 3.1 Performance Benchmarking

In the single-forcing experiment, the diffusion-based hourly model (h-Diffusion) slightly outperforms state-of-the-art MTS-LSTM and MF-LSTM (Figure 2c). The diffusion-based model h-Diffusion achieves a median hourly-scale NSE of 0.774, slightly higher than MTS-LSTM (0.763) and MF-LSTM (0.756). This moderate improvement when compared to already elite alternatives may be related to diffusion architecture's ability to incorporate daily simulated hydrologic states as prior information, which MTS-LSTM or MF-LSTM does not do. h-Diffusion performs best in the western US (Figure 2a), where catchments have large hydro-climatic diversity, including humid, arid, snow-dominated, and complex-terrain basins. Above 0.6 NSE values occur across the eastern half CONUS, which features flatter topography and humid conditions. The daily δHBV model also performs well in this region. h-Diffusion consistently refines the daily simulations widely across the USA, except in the Northern Great Plains, central Texas and scattered basins along the Appalachian front (Figure 2b). The relative weakness of h-Diffusion in the Appalachian front range of the eastern US compared to the daily model (Figure 2b) suggests hourly precipitation in the region may be difficult to capture. Lower performance along the Great Plains, similar to those reported by Gauch et al. (2021) and even lower than the daily model, could be related to large catchment areas in this region and uneven precipitation, which could be improved using multiscale-trained distributed differentiable models in Song et al., (2025).

The spatial patterns of NSEs are quite similar between single- and multiple-forcing h-Diffusion. In the multiple-forcing experiment, the median NSEs of h-Diffusion (0.804) and MF-LSTM (0.805) are nearly identical, and both are slightly below that of MTS-LSTM (0.812) (Figure 2d). All models show clear performance gains compared with their single-forcing counterparts, indicating that they can effectively leverage the richer meteorological information from the multiple forcing products (Daymet, Maurer, and NLDAS) to improve streamflow predictions. We note, however, the advantage of multiforcing is a hypothetical one that needs further demonstration in a forecasted forcing (rather than reanalysis) scenario.

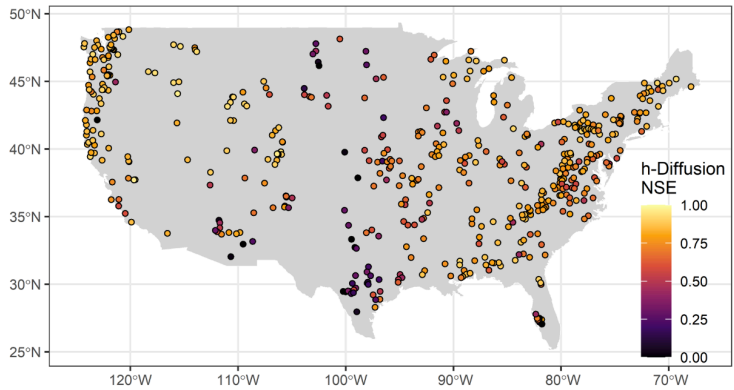
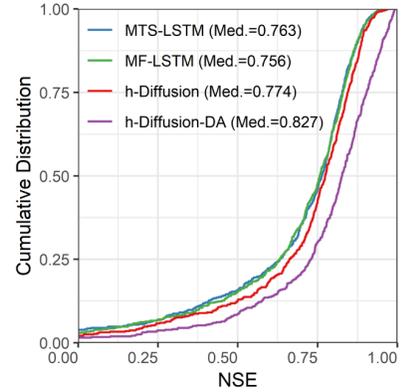
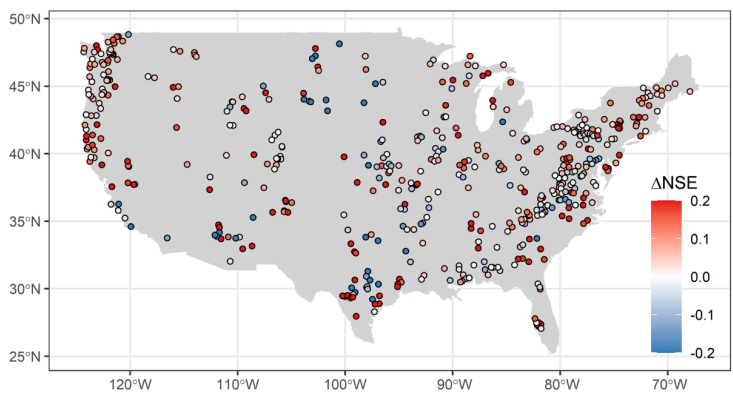
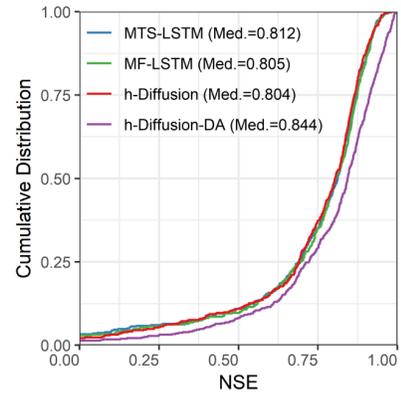

*Figure 2. Nash–Sutcliffe Efficiency (NSE) comparison. (a) Map of NSE by basin for the single-forcing h-Diffusion model. (b) Difference between hourly-scale NSE of h-Diffusion and daily-scale NSE of daily δHBV. (c) Cumulative NSE distributions of different single-forcing models. (d) Cumulative NSE distributions of different multiple-forcing models.*

We next examine the improvements of probabilistic skill of the diffusion-based models in high-flow prediction relative to the baseline methods. The single-forcing h-Diffusion shows consistently positive median CRPSS values (see Equation 9 in Methods, >0 means performance improvement over the reference model) compared with MTS-LSTM across all high-flow percentiles (Figure 3), with $p < 0.05$ in all Wilcoxon signed-rank tests. The median CRPSS ranges from 0.02 to 0.08. When compared with MF-LSTM, h-Diffusion also shows overall positive CRPSS values, although the improvements are not statistically significant for the most extreme events (top 0.2% and 0.1% flows). For the h-Diffusion-DA, the overall probabilistic skill improves much more noticeably. Although the CRPSS for the top 0.1% of flows is not significantly greater than zero, the model shows higher skill for less extreme events. The median CRPSS increases from 0.06 for the top 0.2 % flows to 0.34 for the top 10% flows. These results suggest that diffusion-based models have higher probabilistic prediction skills, especially for ordinary high flows, while the benefit becomes less pronounced for rare extreme floods, possibly due to limited sample size that affects both model learning and statistical tests.

The multi-forcing configuration shows smaller differences than the single-forcing case, with fewer statistically significant median CRPSS values greater than zero across all high-flow percentiles (Figure 3b). For h-Diffusion, none of the median CRPSS values are significantly greater than zero across high-flow percentiles. Nevertheless, h-Diffusion-DA improves prediction skill up to the top 1% flow percentile, with median CRPSS values rising from 0.04 (top 1 %) to 0.27 (top 10 %). A possible explanation is that the larger extreme events are not strongly related to 5-hour-ahead discharges and thus are not significantly improved by inpainting. Another explanation is that the statistical measure of CRPS becomes less reliable for lower-frequency events due to limited sample sizes. Overall, these results demonstrate that the stochasticity in diffusion models can be leveraged to provide accurate probabilistic hourly streamflow predictions.

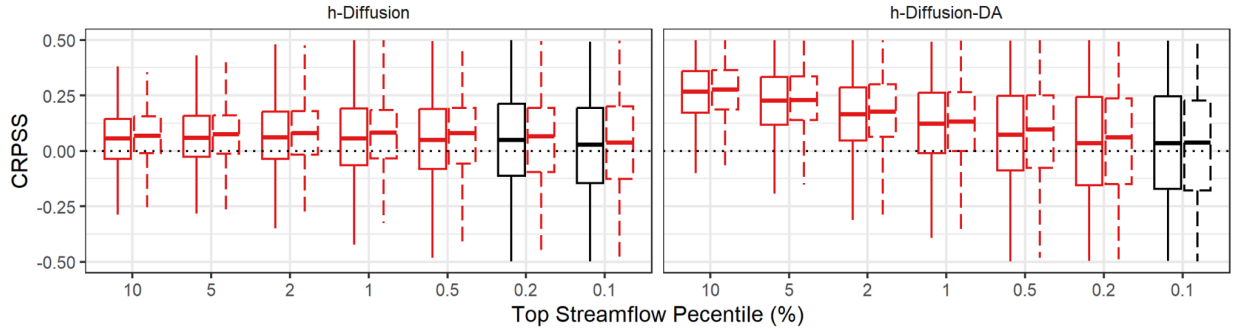
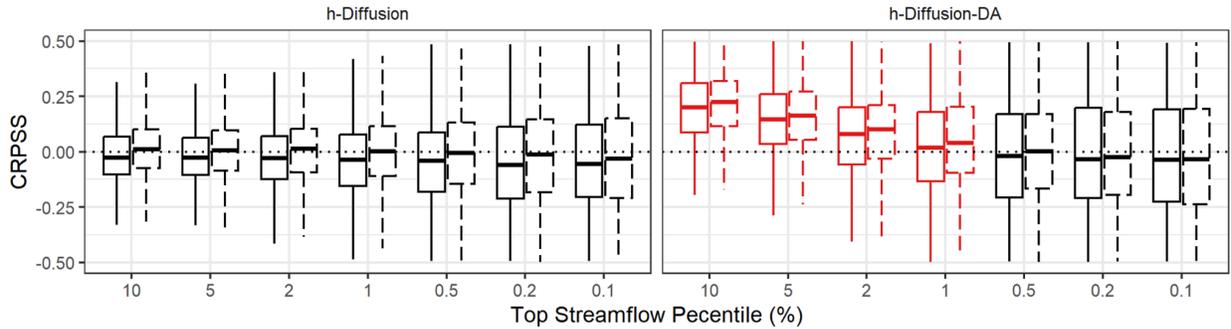

*Figure 3. Continuous Ranked Probability Skill Score (CRPSS) of h-Diffusion and h-Diffusion-DA over MF-LSTM and MTS-LSTM across top percentiles of peak flow for all basins. Positive CRPSS values indicate improved probabilistic prediction skill compared with the reference model. The statistical test is the Wilcoxon signed-rank test for sample median > 0.*

3.2 The Effect of Inpainting for Training-Free Data Assimilation

We now examine how inpainting further contributes to performance gains compared with h-Diffusion, assuming that we have a five-hour lead time in obtaining observations before the 24-hour prediction window. Figure 4 compares h-Diffusion-DA with the baseline h-Diffusion under the single-forcing configuration. As shown earlier in Figure 2c, the median NSE increases from 0.774 to 0.827 when applying inpainting. Figure 4(a) further shows that the largest NSE improvements (> 0.2) occur in the central region, the southeastern coast, and the western arid and snow-dominated basins. These areas typically either exhibit lower baseline performance or complex runoff generation mechanisms. The use of short-term observed streamflow information helps constrain these uncertainties, which leads to improved predictive skill. In contrast, humid basins in the eastern US and western coast, where forcing–streamflow relationships are more stable, show smaller gains.

For high-flow performance, the inpainting model also shows consistent advantages, as evaluated by the CRPSS metric, which quantifies the improvement of h-Diffusion-DA over the reference model (h-Diffusion) in probabilistic prediction skill. The median CRPSS reaches 0.08 for the top 1% flows and rises to 0.26 for the top 10% flows, again suggesting that the assimilation of recent observations benefits more frequent high-flow events. Similar to NSE, the spatial pattern of CRPSS improvement (Figure 4b) shows high CRPSS values ($> 0.5$) observed in the western arid and snow regions and the southeastern coastal basins. Differently, basins in the central plains show limited improvement, while basins around the Great Lakes region show positive gains.

Figure 4(c) illustrates the time series of observations, h-Diffusion, and h-Diffusion-DA predictions for four representative flood events (all above the top 1% threshold). In those events, the baseline h-Diffusion consistently underestimates flood peaks. However, h-Diffusion-DA not only provides more accurate ensemble-mean estimates but also narrower uncertainty ranges, which combine to result in significantly lower last-24-hour average CRPS values. The effectiveness of inpainting arises from both the early underestimation signals and the sequential consistency of the diffusion generation process: given more recent observations, the model re-aligns the generation toward regimes consistent with these observations, thus shrinking the potential uncertainty envelope.

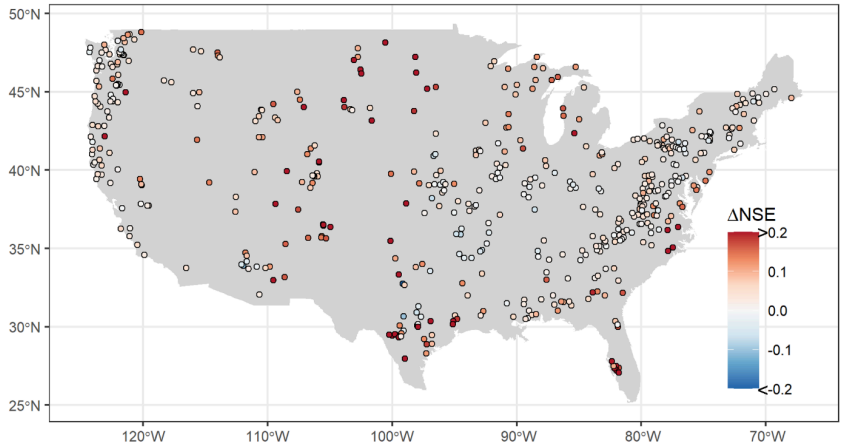
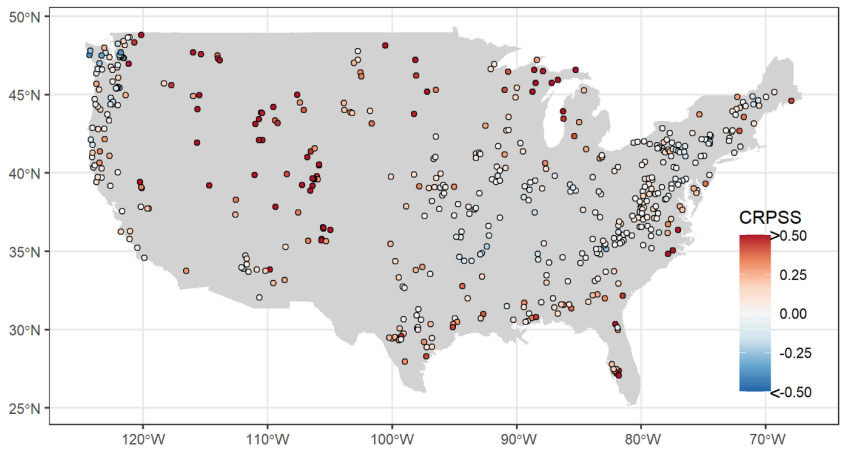
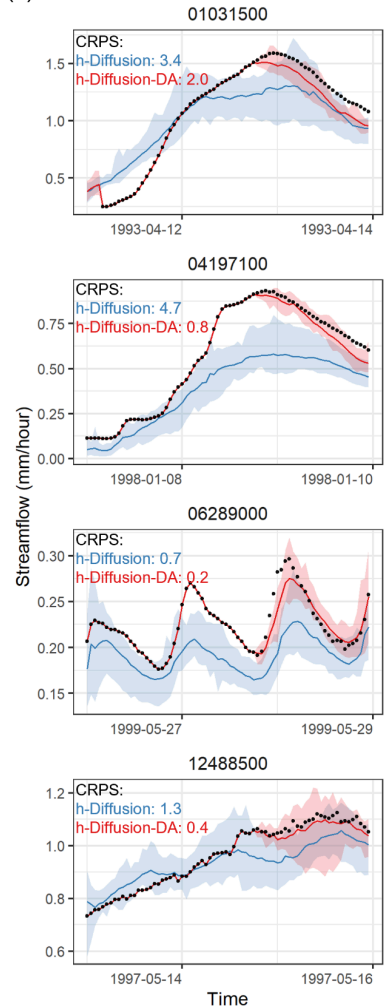

*Figure 4. The effect of inpainting. (b) Nash–Sutcliffe Efficiency (NSE) difference between h-Diffusion-DA and h-Diffusion across top 1% peak flow. (b) Continuous Ranked Probability Skill Score (CRPSS) of h-Diffusion-DA over h-Diffusion across top 1% peak flow. (c) Illustration of how inpainting improves probabilistic streamflow predictions on the last 24 hours after assimilating 5-hour-ahead steamflow observations. All results in the figure are based on single-forcing configuration.*

3.3 Unique Features of the Diffusion Model and Broader Applicability

**Unified Daily–Hourly Downscaling and Correction Framework**
A feature of the proposed diffusion-based model is its capability to perform either daily streamflow downscaling and correction within a unified framework (Mardani et al., 2025). This capability comes from the mixed-sample training strategy, where either observed or simulated daily streamflow can serve as conditional input, and a flag variable switches between the two operation modes. This strategy allows the model to separately learn two components: the low-resolution magnitudes represented by daily streamflow and the high-frequency dynamics represented by hourly fluctuations. For example, using the single-forcing h-Diffusion model with pure daily discharge observations as input yields a median NSE of 0.966, representing the upper performance limit if a perfect daily hydrologic model were available. The high downscaling performance also suggests that the hourly dynamics are well captured, and further improvement largely depends on the potential of improving the daily simulation.

The model's ability to learn both components comes from the denoising multi-step inference, which can be viewed as solving a series of subtasks (Ou et al., 2025). As shown in Figure 5, the model performs 50 denoising steps to gradually refine predictions. If the conditioning daily streamflow is accurate (Figure 5a), the model focuses on reconstructing hourly variations. When the daily input systematically underestimates streamflow (Figure 5b), the denoising process simultaneously produces realistic hourly variations and corrects the magnitude, i.e., conducting downscaling and bias correction jointly within the 50-step generation. This feature enables the model to bridge the gap between daily and sub-daily hydrologic predictions.

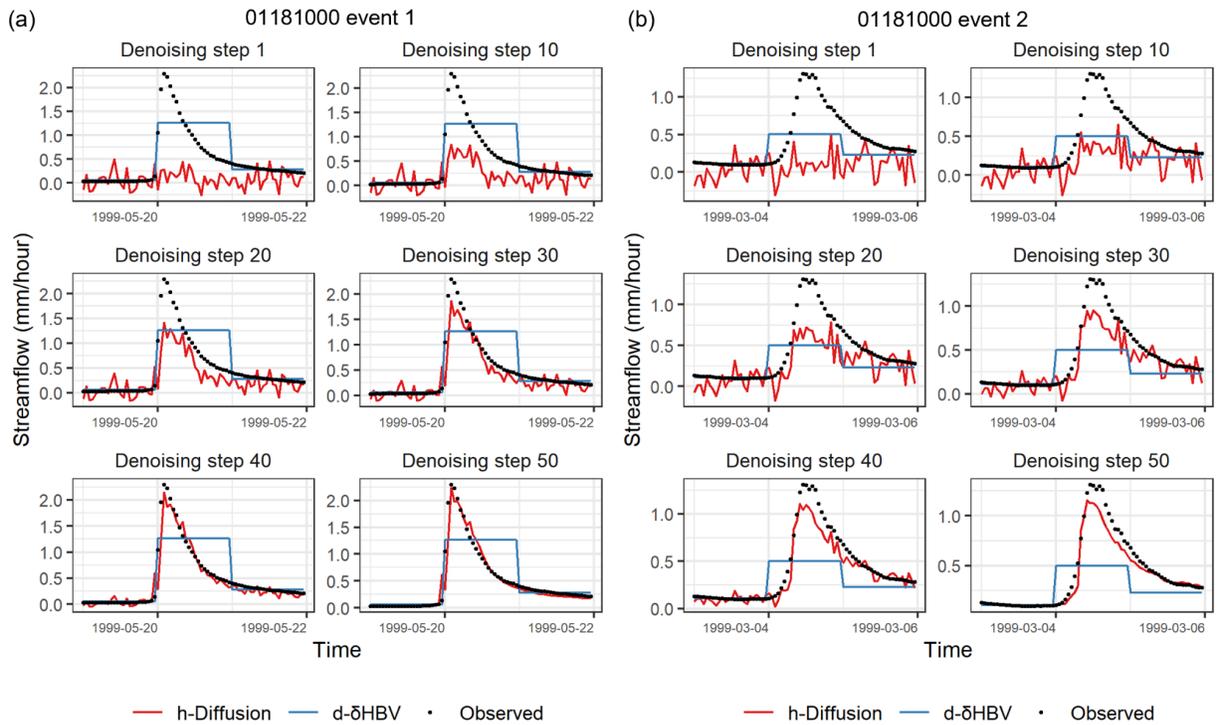

*Figure 5. Denoising processes of streamflow prediction for gauge 01181000 during (a) 1999 May event and (b) 1999 March event.*

**Probabilistic Representation**

The second advantage of the proposed model, as discussed in Ou et al. (2025), is its inherent ability for probabilistic forecasts. Deterministic models such as MTS-LSTM and MF-LSTM require training multiple neural networks with different random seeds to account for prediction uncertainty. However, diffusion-based models learn an explicit stochastic mapping from inputs to the conditional distribution of streamflow. Once trained, the model can generate any number of realizations, each representing a statistical sampling from the learned conditional distribution. Data assimilation via inpainting can further reduce the uncertainty envelope, which even using an ensemble of MTS-LSTM models could not achieve. When presented with 5-hour-ahead discharge observations, it can shrink the forecast uncertainty going forward (Figure 4). This feature provides a clearer interpretation of uncertainty: it reflects the inherently unknowable variability of the hydrologic process (Chao et al., 2025), where multiple possible realizations exist given the same input forcing and attributes. Instead, uncertainty from training random seeds partly represents the randomness in the training procedure of machine learning such as initialization or optimization noise.

Moreover, the diffusion-based models provide ensemble prediction with only one trained model, reducing the computational cost compared to training multiple deterministic models. Using a single NVIDIA GeForce RTX 3090 Ti GPU, training the h-Diffusion model on the CAMELS-US dataset for the 13-year period (1990–2003) takes approximately 7 hours. Generating 10 realizations of hourly streamflow predictions for all 516 basins over a 5-year period (2009–2014) takes about 2.5 hours, and producing 10 realizations with inpainting (h-Diffusion-DA) for another 5-year period (2003–2008) takes around 12 hours.

**Generic Training-or-Optimization-Free Data Integration through Inpainting**

Diffusion-based models support flexible data integration inherently attributed to the inpainting technique. In this technique, a portion of the predicted streamflow is replaced by available recent observations, and the model reconstructs the remaining part through the same denoising steps. This feature provides seamless incorporation of real-time data without needing an independent data assimilation module, e.g., those described in (Jamaat et al., 2025; Nearing et al., 2022b), with additional optimization effort. The results demonstrate the effectiveness of the technique: the inpainting model h-Diffusion-DA achieves the highest performance, with NSE = 0.827 under the single-forcing configuration and NSE = 0.844 under the multiple-forcing configuration (Figures 2b and 2d). It significantly improves hourly streamflow modeling in hard-to-predict regions, such as central and southeastern coastal basins (Figure 4a). In addition, inpainting is an optional component during model inference, meaning it can be applied in gauged catchments where recent observations are available, and disabled in ungauged catchments. In contrast, conventional data-integration models (Fang & Shen, 2020; Feng et al., 2020; Nearing et al., 2022b; Y. Song et al., 2024) can only be applied to gauged basins because they need systematic training data. This flexibility makes the proposed diffusion-based models suitable for large-scale hydrologic applications, where most rivers are ungauged but the same model can operate across all basins.

Note here this work focuses on the hourly prediction, with only a limited demonstration of the data assimilation capability. Using a generative model, the conditionally sampled states based on observations should maintain internal consistency compared to that from a variational optimization. This would allow better inverse modeling for states and forcings. Furthermore, the fact it can capture high-dimensional joint distribution should in theory provide benefits for extreme values compared to linear assumptions used in Ensemble-Kalman filtering. However, these features may or may not necessarily lead to gains in forecast performance. To understand these features along with computational efficiency issues warrants a deeper investigation in a separate study.

## 4. Conclusions

Hourly streamflow modeling provides critical high-resolution information for flood forecasting and water resource management. In this study, we developed h-Diffusion, a diffusion-based stochastic model for hourly streamflow prediction. The model flexibly incorporates daily streamflow information either from observations or hydrologic simulations, making it physically informed and applicable to both daily-to-hourly downscaling and direct hourly prediction. By integrating the inpainting technique, the augmented model h-Diffusion-DA can assimilate recent observations directly during inference without retraining or additional optimization.

Across 516 CAMELS-US basins, h-Diffusion and h-Diffusion-DA outperform state-of-the-art data-driven baselines (MTS-LSTM and MF-LSTM). Under the single-forcing configuration, h-Diffusion improves the median NSE to 0.774, and inpainting further raises it to 0.827; under multiple-forcing settings, h-Diffusion-DA achieves the best performance with NSE = 0.844. Both diffusion-based models also improve probabilistic prediction skill over baselines, with significantly positive CRPSS values for frequent high-flow events (top 10–1%). The inpainting approach provides notable gains in low-performing and hydrologically complex regions.

In summary, diffusion-based models offer a unified probabilistic framework for streamflow prediction, downscaling, and data assimilation. Their flexibility and built-in uncertainty representation make them a promising method for operational hydrologic forecasting. Future work should focus on computational optimization and integration with physical process models to further unlock their potential.

## Data availability

The hourly NLDAS forcing and the hourly streamflow can be downloaded at https://zenodo.org/records/4072701 (Gauch et al., 2020). The CAMELS-US dataset is from Addor et al. (2017).

## Code availability

The code will be made available on GitHub and Zenodo.

# References

Acuña Espinoza, E., Kratzert, F., Klotz, D., Gauch, M., Álvarez Chaves, M., Loritz, R., & Ehret, U. (2025). Technical note: An approach for handling multiple temporal frequencies with different input dimensions using a single LSTM cell. *Hydrology and Earth System Sciences*, *29*(6), 1749–1758. https://doi.org/10.5194/hess-29-1749-2025

Addor, N., Newman, A. J., Mizukami, N., & Clark, M. P. (2017). The CAMELS data set: catchment attributes and meteorology for large-sample studies. *Hydrol. Earth Syst. Sci.*

Alfieri, L., Pappenberger, F., Wetterhall, F., Haiden, T., Richardson, D., & Salamon, P. (2014). Evaluation of ensemble streamflow predictions in Europe. *Journal of Hydrology*, *517*, 913–922. https://doi.org/10.1016/j.jhydrol.2014.06.035

Chao, J., Pan, B., Chen, Q., Yang, S., Wang, J., Nai, C., et al. (2025). Learning to Infer Weather States Using Partial Observations. *Journal of Geophysical Research: Machine Learning and Computation*, *2*(1), e2024JH000260. https://doi.org/10.1029/2024JH000260

Dai, T.-Y., & Ushijima-Mwesigwa, H. (2025). PrecipDiff: Leveraging Image Diffusion Models to Enhance Satellite-Based Precipitation Observations. *Proceedings of the AAAI Conference on Artificial Intelligence*, *39*(27), 27932–27939. https://doi.org/10.1609/aaai.v39i27.35010

Fang, K., & Shen, C. (2020). Near-Real-Time Forecast of Satellite-Based Soil Moisture Using Long Short-Term Memory with an Adaptive Data Integration Kernel. *Journal of Hydrometeorology*, *21*(3), 399–413. https://doi.org/10.1175/JHM-D-19-0169.1

Feng, D., Fang, K., & Shen, C. (2020). Enhancing Streamflow Forecast and Extracting Insights Using Long‑Short Term Memory Networks With Data Integration at Continental Scales. *Water Resources Research*, *56*(9), e2019WR026793. https://doi.org/10.1029/2019WR026793

Feng, D., Liu, J., Lawson, K., & Shen, C. (2022). Differentiable, Learnable, Regionalized Process‑Based Models With Multiphysical Outputs can Approach State‑Of‑The‑Art Hydrologic Prediction Accuracy. *Water Resources Research*, *58*(10), e2022WR032404. https://doi.org/10.1029/2022WR032404

Feng, D., Beck, H., Lawson, K., & Shen, C. (2023). The suitability of differentiable, physics-informed machine learning hydrologic models for ungauged regions and climate change impact assessment. *Hydrology and Earth System Sciences*, *27*(12), 2357–2373. https://doi.org/10.5194/hess-27-2357-2023

Furl, C., Sharif, H., Zeitler, J. W., Hassan, A. E., & Joseph, J. (2018). Hydrometeorology of the catastrophic Blanco river flood in South Texas, May 2015. *Journal of Hydrology: Regional Studies*, *15*, 90–104. https://doi.org/10.1016/j.ejrh.2017.12.001

Gauch, M., Kratzert, F., Klotz, D., Nearing, G., Lin, J., & Hochreiter, S. (2020). Models and Predictions for "Rainfall-Runoff Prediction at Multiple Timescales with a Single Long Short-Term Memory Network" [Data set]. Zenodo. https://doi.org/10.5281/ZENODO.4071885

Gauch, M., Kratzert, F., Klotz, D., Nearing, G., Lin, J., & Hochreiter, S. (2021). Rainfall–runoff prediction at multiple timescales with a single Long Short-Term Memory network. *Hydrology and Earth System Sciences*, *25*(4), 2045–2062. https://doi.org/10.5194/hess-25-2045-2021

Georgakakos, K. P., Modrick, T. M., Shamir, E., Campbell, R., Cheng, Z., Jubach, R., et al. (2022). The Flash Flood Guidance System Implementation Worldwide: A Successful Multidecadal Research-to-Operations Effort. *Bulletin of the American Meteorological Society*, *103*(3), E665–E679. https://doi.org/10.1175/BAMS-D-20-0241.1

Gneiting, T., & Raftery, A. E. (2007). Strictly Proper Scoring Rules, Prediction, and Estimation. *Journal of the American Statistical Association*, *102*(477), 359–378. https://doi.org/10.1198/016214506000001437

Hellin, J., Haigh, M., & Marks, F. (1999). Rainfall characteristics of hurricane Mitch. *Nature*, *399*(6734), 316–316. https://doi.org/10.1038/20577

Ho, J., Jain, A., & Abbeel, P. (2020, December 16). Denoising Diffusion Probabilistic Models. arXiv. https://doi.org/10.48550/arXiv.2006.11239

Jamaat, A., Song, Y., Rahmani, F., Liu, J., Lawson, K., & Shen, C. (2025). Update hydrological states or meteorological forcings? Comparing data assimilation methods for differentiable hydrologic models. *Journal of Hydrology*, *663*, 134137. https://doi.org/10.1016/j.jhydrol.2025.134137

Lugmayr, A., Danelljan, M., Romero, A., Yu, F., Timofte, R., & Gool, L. V. (2022, August 31). RePaint: Inpainting using Denoising Diffusion Probabilistic Models. arXiv. https://doi.org/10.48550/arXiv.2201.09865

Ma, X., Dong, X., Tarrant, A., Yang, L., Kotamarthi, R., Wang, J., et al. (2025). Diffusion-Based, Data-Assimilation-Enabled Super-Resolution of Hub-height Winds (Version 1). arXiv. https://doi.org/10.48550/ARXIV.2510.03364

Mardani, M., Brenowitz, N., Cohen, Y., Pathak, J., Chen, C.-Y., Liu, C.-C., et al. (2025). Residual corrective diffusion modeling for km-scale atmospheric downscaling. *Communications Earth & Environment*, *6*(1), 124. https://doi.org/10.1038/s43247-025-02042-5

Nash, J. E., & Sutcliffe, J. V. (1970). River flow forecasting through conceptual models part I — A discussion of principles. *Journal of Hydrology*, *10*(3), 282–290. https://doi.org/10.1016/0022-1694(70)90255-6

National Severe Storms Laboratory. (n.d.). Flood Basics [text]. Retrieved October 9, 2025, from https://www.nssl.noaa.gov/education/svrwx101/floods/

Nearing, G. S., Klotz, D., Frame, J. M., Gauch, M., Gilon, O., Kratzert, F., et al. (2022a). Technical note: Data assimilation and autoregression for using near-real-time streamflow observations in long short-term memory networks. *Hydrology and Earth System Sciences*, *26*(21), 5493–5513. https://doi.org/10.5194/hess-26-5493-2022

Nearing, G. S., Klotz, D., Frame, J. M., Gauch, M., Gilon, O., Kratzert, F., et al. (2022b). Technical note: Data assimilation and autoregression for using near-real-time streamflow observations in long short-term memory networks. *Hydrology and Earth System Sciences*, *26*(21), 5493–5513. https://doi.org/10.5194/hess-26-5493-2022

Ou, Z., Nai, C., Pan, B., Zheng, Y., Shen, C., Jiang, P., et al. (2025). Probabilistic diffusion models advance extreme flood forecasting. *Geophysical Research Letters*, *52*(15), e2025GL115705. https://doi.org/10.1029/2025GL115705

Song, J., Meng, C., & Ermon, S. (2022, October 5). Denoising Diffusion Implicit Models. arXiv. https://doi.org/10.48550/arXiv.2010.02502

Song, Y., Tsai, W.-P., Gluck, J., Rhoades, A., Zarzycki, C., McCrary, R., et al. (2024). LSTM-Based Data Integration to Improve Snow Water Equivalent Prediction and Diagnose Error Sources. *Journal of Hydrometeorology*, *25*(1), 223–237. https://doi.org/10.1175/JHM-D-22-0220.1

Song, Y., Sawadekar, K., Frame, J. M., Pan, M., Clark, M., Knoben, W. J. M., et al. (2025). Physics-informed, differentiable hydrologic models for capturing unseen extreme events. ESS Open Archive. https://doi.org/10.22541/essoar.172304428.82707157/v2

World Meteorological Organization. (2020, March 23). Flash Flood Guidance System: Response to one of the deadliest hazards. Retrieved October 9, 2025, from https://wmo.int/media/magazine-article/flash-flood-guidance-system-response-one-of-deadliest-hazards

Xia, Y., Mitchell, K., Ek, M., Sheffield, J., Cosgrove, B., Wood, E., et al. (2012). Continental-scale water and energy flux analysis and validation for the North American Land Data Assimilation System project phase 2 (NLDAS-2): 1. Intercomparison and application of model products. *Journal of Geophysical Research: Atmospheres*, *117*(D3). https://doi.org/10.1029/2011JD016048

Yi, C., Yu, M., Qian, W., Wen, Y., & Yang, H. (2025). Efficient Kilometer-Scale Precipitation Downscaling with Conditional Wavelet Diffusion (Version 1). arXiv. https://doi.org/10.48550/ARXIV.2507.01354